\documentclass[12pt]{article}
\usepackage{a4}
\usepackage{amsmath}
\usepackage{amssymb}
\usepackage[dvips]{graphicx}
\usepackage{subfigure}
\def\beq{\begin{equation}}
\def\eeq{\end{equation}}
\def\beqa{\begin{eqnarray}}
\def\eeqa{\end{eqnarray}}
\def\n{\nonumber}

\newcommand{\bel}{\begin{equation}\label}

\newcommand {\bc}{\begin{center}}
\newcommand {\ec}{\end{center}}
\newcommand {\tr}{{\rm tr}\,}

\def\dag{\dagger}

\def\vs5{\vspace*{5mm}}
\def\vs1{\vspace*{1cm}}
\def\vs2{\vspace*{2cm}}
\def\hs5{\vspace*{5mm}}
\def\hs1{\hspace*{1cm}}
\def\hs2{\hspace*{2cm}}
\def\vs50{\vspace*{50mm}}
\def\vs20{\vspace*{20mm}}
\def\tr{\hbox{tr}}

\begin{document}
%%%%%%%%%%%%%%%%%%%%%%%%%%%%%%%%%%%%%%%%%%%%%%%%
 
\begin{flushright}
{SAGA-HE-218}\\
{KEK-TH-1000}
\end{flushright}
\vskip 0.5 truecm
%%%%%%%%%%%%%%%%%%%%%%%%%%%%%%%%%%%%%%%%%%%%%%%%
 
%%%%%%%%%%%%%%%%%%%%%%%%%%%%%%%%%%%%%%%%%%%%%%%%
 
\begin{center}
{\Large{\bf Dynamical generation of a nontrivial index\\ 
on the fuzzy 2-sphere}}
\vskip 1.0cm
 
{\large Hajime Aoki$^{a}$\footnote{e-mail
 address: haoki@cc.saga-u.ac.jp},
 Satoshi Iso$^b$\footnote{e-mail
 address: satoshi.iso@kek.jp},
 Toshiharu Maeda$^{a,b}$\footnote{e-mail
 address: 03td23@edu.cc.saga-u.ac.jp, maeda@post.kek.jp}\\
 and \\
 Keiichi Nagao$^b$\footnote{e-mail
 address: nagao@post.kek.jp}}
\vskip 0.5cm

$^a${\it Department of Physics, Saga University, Saga 840-8502,
Japan  }\\
 
%{\it and }\\
 
$^b${\it Institute of Particle and Nuclear Studies, \\
High Energy Accelerator Research Organization (KEK)\\
Tsukuba 305-0801, Japan}
 
\end{center}
 
\vskip 1cm
\begin{center}
\begin{bf}
Abstract
\end{bf}
\end{center}
In the previous paper \cite{AIN3} we studied 
the 't Hooft-Polyakov (TP) monopole configuration 
in the $U(2)$ gauge theory on the fuzzy 2-sphere 
and showed that it has a nonzero topological charge 
in the formalism based on the Ginsparg-Wilson relation.
In this paper, by showing that 
the TP monopole configuration is stabler than
the $U(2)$ gauge theory without any condensation
in the  Yang-Mills-Chern-Simons matrix model,
we will present a mechanism for dynamical generation
of a nontrivial index.
We further analyze the instability and decay processes of 
the  $U(2)$ gauge theory
and the TP monopole configuration.

%%%%%%%%%%%%%%%%%%%%%%%%%%%%%%%%%%%%%%%%%%%%%%%% 
\newpage
\setcounter{footnote}{0}
\section{Introduction}
\setcounter{equation}{0}

Matrix models are promising candidates to 
formulate the superstring theory nonperturbatively. 
In IIB matrix model\cite{IKKT}, 
which is the 
dimensionally reduced model of the 10-dimensional super 
Yang-Mills theory, both the space-time and the matter are 
described in terms of matrices, and noncommutative (NC) 
geometries\cite{Connes} naturally appear\cite{CDS,NCMM}. 

One of the important subjects of the matrix model 
is a construction of 
configurations with nontrivial indices in finite NC geometries.  
It is not only interesting 
from the mathematical point of view 
but necessary from the physical requirement. 
Namely, compactifications of extra dimensions
with nontrivial indices can realize 
chiral gauge theories in our space-time. 
Nontrivial configurations in NC geometries have been constructed 
by using algebraic K-theory and projective modules 
\cite{non-trivial_config,non_chi,Valtancoli:2001gx,
Steinacker:2003sd,Karabali:2001te,Carow-Watamura:2004ct}.
However the relation to indices of Dirac operators 
are not clear in these formulations.

The formulation of NC geometries
in Connes' prescription is based on the spectral
triple (${\cal A}$,${\cal H}$,${\cal D}$), where a chirality
operator and a Dirac operator which anti-commute are
introduced\cite{Connes}. 
Since NC geometries on compact manifolds have only 
finite degrees of freedom, 
a more suitable framework to discuss the problems mentioned 
above will be to modify the Connes' spectral triple so that 
the chirality operator and the Dirac operator satisfy the
Ginsparg-Wilson (GW) relation\cite{GinspargWilson}. 
GW relation has been developed in the lattice gauge theory. 
The exact chiral symmetry\cite{Luscher,Nieder} and 
the index theorem\cite{Hasenfratzindex,Luscher} 
at a finite cutoff can be realized 
by using the GW Dirac operator\cite{Neuberger}.

In ref.\cite{AIN2}, we have provided 
a general prescription to construct chirality operators and 
Dirac operators satisfying the GW relation 
in general gauge field backgrounds on finite NC geometries.
As a concrete example 
we considered the fuzzy 2-sphere\cite{Madore}.
%%%%%%%%%%%%%%
On the fuzzy 2-sphere two types of Dirac operators, 
$D_{\rm WW}$\cite{Carow-Watamura:1996wg} and
$D_{\rm GKP}$\cite{Grosse:1994ed,IKTW}, had been constructed. 
$D_{\rm WW}$ has doublers and the correct chiral anomaly 
cannot be reproduced. 
On the other hand, $D_{\rm GKP}$ breaks 
chiral symmetry at finite matrix size, 
and the chiral structures are not 
clear,
though the chiral anomaly 
can be reproduced correctly in the commutative 
limit\cite{chiral_anomaly,non_chi,chiral_anomaly2,AIN1}.
%%%%%%%%%%
Hence the formalism based on the GW relations is 
more suitable to study chiral structures on the fuzzy 2-sphere.
The GW Dirac operator in vanishing gauge field 
was constructed in \cite{balagovi,balaGW}.
In ref.\cite{AIN2}, we constructed GW Dirac operator in 
general gauge field configurations.
Owing to the GW relation, an index theorem can be proved 
even for finite NC geometries.
We have defined a topological charge,
and showed that it takes only integer values, 
and becomes the Chern character in the commutative limit
\cite{AIN2,Ydri:2002nt,Balachandran:2003ay,AIN3}. \footnote{
The GW relation can be implemented also on the NC 
torus by using the Neuberger's GW Dirac 
operator\cite{Nishimura:2001dq}.  
The correct chiral anomaly was 
reproduced in \cite{Iso:2002jc} by using a topological 
method in \cite{Fujiwara:2002xh}. 
The correct parity anomaly was also reproduced in 
\cite{Nishimura:2002hw}.}

We then studied the 't Hooft-Polyakov (TP) monopole configuration
as a topologically nontrivial configuration
in \cite{AIN3}.
We showed that this configuration 
is a NC analogue of the commutative TP monopole 
by explicitly studying the form of the configuration.
We then redefined the topological charge 
by inserting a projection operator,
and showed that it reproduces the correct form of 
the topological charge in the commutative limit,
namely, the magnetic flux for the unbroken $U(1)$ component.
We also showed that the value of the topological charge 
is one for the TP monopole configuration. 
Therefore, if the $U(2)$ gauge theory on the fuzzy 2-sphere
decays to the TP monopole configuration, 
it means a nontrivial index is generated dynamically.
It can also be interpreted as a spontaneous symmetry breaking:
$U(2) \rightarrow U(1) \times U(1)$ on the fuzzy 2-sphere. 

In this paper, by showing that 
the TP monopole configuration 
is stabler than the $U(2)$ gauge theory 
without condensation
in the Yang-Mills-Chern-Simons matrix model
\cite{Alekseev:2000fd}\cite{IKTW}, 
we will present a mechanism of the 
dynamical generation of an index through 
the spontaneous symmetry breaking.
In this matrix model,
the $U(2)$ gauge theory on the fuzzy 2-sphere
corresponds to the 2-coincident
fuzzy 2-spheres.
The TP monopole configuration is  equivalent to the 
2-concentric fuzzy 2-spheres 
whose matrix sizes are different by two. 
More general 2-concentric fuzzy 2-spheres 
whose matrix sizes are different by integers
$2m$ correspond to a monopole configuration 
with the topological charge $-|m|$.
We will also discuss these general monopole configurations.

The stability of these configurations are
investigated  in papers 
\cite{Alekseev:2000fd,IKTW,Bal:2001cs,Valtancoli:2002rx,Imai:2003vr,Azuma:2004zq,Castro-Villarreal:2004vh,Azuma:2004qe,Azuma:2004ie}.
Especially $k$-coincident fuzzy spheres are analyzed 
in \cite{Azuma:2004zq}. 
The configuration gives the $U(k)$ gauge theory 
on the fuzzy sphere. 
In \cite{Azuma:2004zq} the dynamics was studied 
both analytically and 
numerically, and the  $U(k)$ gauge theory has been 
shown to be unstable. 
It decays,
and finally arrives at 
the single fuzzy sphere configuration,
the $U(1)$ gauge theory on the fuzzy sphere.

In this paper,
by studying free energies for the 2-concentric fuzzy spheres 
whose matrix sizes are different by arbitrary integers $2m$,
we investigate how 2-coincident fuzzy spheres decay to 
the single fuzzy sphere.
The configuration with the matrix-size-difference $2(m+1)$ 
is stabler than the one with 
$2m$ at the 
classical level,
and quantum corrections do not change this property.
Thus a configuration of the 2-coincident spheres ($m=0$) 
decays into the TP monopole configuration of $m=1$. 
By repeating such transitions,
it cascades into the single fuzzy sphere configuration.

We further study the decay modes around the configuration of 
the 2-coincident spheres ($m=0$)
and the TP monopole configuration ($m=1$) 
by calculating effective actions along the directions 
of the collective (zero) modes.
The case for $m=0$ receives subtle quantum corrections:
The zero-mode directions become unstable for 
small values of the total matrix size $N$,
and metastable for large N,
but the metastability becomes negligible at large $N$ limit.
The case for $m=1$ is metastable at the classical level,
and quantum corrections do not change this property.

In section 2 we briefly review 
how to define the Dirac operator and the chirality operator
on the fuzzy 2-sphere,
where the GW relation and thus the index theorem
are satisfied naturally.
Then we review the monopole configurations
with the nontrivial topological charges.
In section 3,
we introduce the Yang-Mills-Chern-Simons 
matrix model\cite{Alekseev:2000fd}\cite{IKTW},
and then investigate the instability of 
the monopole configurations in this model. 
We show that for $U(2)$ gauge theory 
the monopole configurations of larger $m$
are stabler,
which means the symmetry is spontaneously broken and 
the index is generated dynamically.
Section 4 is devoted to conclusions and discussions.
In appendix \ref{oneloop_eff} the one-loop effective action 
is calculated for the matrix model.
  
%%%%%%%%%%%%%%%%%%%%%%%%%%%%%%%%%%%%%%%%%%%%%%%% 

\section{Configurations with nontrivial topological charges}
\setcounter{equation}{0}

In this section, after briefly reviewing the ordinary 
Dirac operator on the fuzzy sphere in subsection \ref{sec:f2s},
we see how to construct the Dirac operator and
chirality operator satisfying the GW relation
in subsection \ref{sec:GWindex}. 
We also explain the construction of the topological charge 
and the index theorem on the fuzzy 2-sphere.
Then we explain the monopole configurations with
nontrivial topological charges
in subsection \ref{sec:monopole}.

\subsection{Dirac operator on fuzzy 2-sphere}
\label{sec:f2s}

NC coordinates of the
fuzzy 2-sphere are described by
\begin{equation}
x_i =\alpha L_i, \label{xi}
\end{equation}
where $\alpha$ is the NC parameter,
and $L_i$'s are $n$-dimensional irreducible
representation matrices of $SU(2)$ algebra: 
\begin{eqnarray}
[L_i,L_j]&=&i\epsilon_{ijk}L_k . \label{su2}
\end{eqnarray}
Then we have the following relation,
\begin{eqnarray}
(x_i)^2&=&\alpha^2 \frac{n^2-1}{4}{\bold 1}_n 
\equiv \rho^2 {\bold 1}_n,
\end{eqnarray}
where 
$\rho$ expresses the radius of the fuzzy 2-sphere.
The commutative limit can be taken by $\alpha \to0, n \to \infty$
with $\rho$ fixed.
 
Any wave functions on fuzzy 2-sphere are written as 
$n \times n$ matrices. 
We can expand them in terms of NC
analogues of the spherical harmonics $\hat{Y}_{lm}$, 
which are traceless symmetric products of
the NC coordinates, and has an upper bound for 
the angular momentum $l$ as $l \le n-1$.
Derivatives along the Killing vectors on the sphere 
are written as the adjoint operator of $L_i$ on any hermitian 
matrix $M$:
\begin{eqnarray}
&&{\tilde L}_i M= [L_i, M] =(L_i^L-L_i^R)M 
\label{adjLi}\\
&\leftrightarrow&
{{\cal L}_i}M=-i\epsilon_{ijk}x_j \partial_k M.
\label{comLi}
\end{eqnarray}
In eq.(\ref{adjLi}) the superscript L (R) in $L_i$ means that this
operator acts from the left (right) on matrices. 
Eq.(\ref{comLi}) expresses the commutative limit of 
eq.(\ref{adjLi}).
An integral over the 2-sphere is given by a trace
over matrices:
\begin{equation}
\frac{1}{n} \tr \leftrightarrow
\int \frac{d \Omega}{4 \pi}.
\end{equation}

There are two types of Dirac operators constructed on the fuzzy 
2-sphere \cite{Carow-Watamura:1996wg}\cite{Grosse:1994ed,IKTW}. 
Here we consider the following Dirac operator $D_{\rm GKP}$.  
The fermionic action is defined as
\begin{eqnarray}
S_{\rm GKP}&=& \tr[\bar\Psi D_{\rm GKP} \Psi], \\
D_{\rm GKP}&=& \sigma_i ({\tilde L}_i + \rho a_i ) +1,
\label{DGKP}
\end{eqnarray}
where $a_i$ is a gauge field.  
This action is invariant under the gauge transformation:
\begin{eqnarray}
\Psi &\rightarrow& U \Psi, \n\\
\bar\Psi &\rightarrow& \bar\Psi U^\dagger , \n\\
a_i &\rightarrow& U a_i U^\dag +\frac{1}{\rho} (U L_i U^\dag-L_i),
\label{gaugeTra}
\end{eqnarray}
since a combination
\begin{equation}
A_i \equiv L_i+\rho a_i
\label{defAi}
\end{equation}
transforms covariantly as
\begin{equation}
A_i \rightarrow U A_i U^\dagger. \label{gaugetr_A}
\end{equation}
 
The normal component of $a_i$ to the sphere can
be interpreted as
the scalar field on the sphere.
We define it covariantly as
\begin{equation}
\phi=\frac{1}{n\rho} 
\left( A_i^2 - \frac{n^2-1}{4}
\right)\label{def_NC_phi}. 
\end{equation}
In the commutative limit this scalar field (\ref{def_NC_phi})
becomes $a_i n_i$, where $n_i=x_i/\rho$.
If we take the commutative limit of (\ref{DGKP}),
the Dirac operator $D_{\rm GKP}$ becomes 
\begin{equation}
D_{\rm GKP} \rightarrow 
D_{\rm com}=\sigma_i ({\cal L}_i + \rho a_i ) +1,
\end{equation}
which is the ordinary Dirac operator on the commutative 2-sphere.

Due to the noncommutativity of the coordinates, 
$D_{\rm GKP}$ does not anti-commute with
the chirality operator.
Then, 
by carefully evaluating this nonzero anticommutation relation,
the chiral anomaly 
can be reproduced correctly 
\cite{chiral_anomaly,non_chi,chiral_anomaly2,AIN1}.
However, chiral structures are not transparent 
in this formulation,
and we will define another Dirac operator
suitable for these issues.

\subsection{GW Dirac operator and index theorem}
\label{sec:GWindex}

In order to discuss chiral structures on the fuzzy 2-sphere, 
we define a Dirac operator satisfying the GW 
relation. 
Such a Dirac operator was constructed  
for the free case in \cite{balagovi,balaGW}, 
and for general gauge field configurations in \cite{AIN2}.
According to the formulatoin in \cite{AIN2}, 
we first define two chirality operators:
\begin{eqnarray}
\Gamma^R &=& a\left(\sigma_i L_i^R -\frac{1}{2}\right)
, \label{gammaR}\\
\hat\Gamma&=&\frac{H}{\sqrt{H^2}} \label{gammahat},
\end{eqnarray}
where
\begin{equation}
H=a\left(\sigma_i A_i +\frac{1}{2}\right), 
\end{equation}
and
\begin{equation}
a=\frac{2}{n}
\end{equation}
is introduced as a NC analogue of a lattice-spacing.
These chirality operators satisfy
\begin{equation}
(\Gamma^R)^\dagger=\Gamma^R, \
(\hat\Gamma)^\dagger=\hat\Gamma, \
(\Gamma^R)^2=(\hat\Gamma)^2=1.
\end{equation}

We next define the GW Dirac operator as
\begin{equation}
D_{\rm GW} = -a^{-1}\Gamma^R (1- \Gamma^R \hat{\Gamma}).
\label{defDGW}
\end{equation}
Then the action
\begin{equation}
S_{\rm GW}= \tr [\bar\Psi D_{\rm GW} \Psi]
\end{equation}
is invariant under the gauge transformation
(\ref{gaugeTra}).
If we take the commutative limit, $D_{\rm GW}$ becomes 
\begin{equation}
D_{\rm GW} \rightarrow 
D'_{\rm com}=\sigma_i ({\cal L}_i + \rho P_{ij} a_j ) +1,
\end{equation}
where $P_{ij}=\delta_{ij}-n_i n_j$ 
is the projector to the tangential directions 
on the sphere.

By the definition (\ref{defDGW}),
the GW relation  
\begin{equation}
\Gamma^R D_{\rm GW}+D_{\rm GW} \hat{\Gamma}=0 
\label{GWrelation}
\end{equation}
is satisfied. 
Then we can prove the following index theorem:
\begin{equation}
{\rm{index}}D_{\rm GW}\equiv (n_+ - n_-)=\frac{1}{2}
{\cal T}r(\Gamma^R +\hat{\Gamma}), 
\label{indexth}
\end{equation}
where
$n_{\pm}$ are the numbers of zero eigenstates of $D_{\rm GW}$
with a positive (or negative) chirality (for either $\Gamma^R$
or $\hat{\Gamma}$)
and ${\cal T}r$ is a trace of operators acting on matrices.

The rhs of (\ref{indexth}) have the following properties. 
Firstly, it takes only integer values since both $\Gamma^R$ 
and $\hat{\Gamma}$ have a form of sign operator 
by the definitions (\ref{gammaR}), (\ref{gammahat}). 
Secondly it becomes the Chern character on the 2-sphere 
in the commutative limit. 
Finally,
in order to discuss a topological charge 
for topologically nontrivial configurations in a spontaneously 
symmetry broken gauge theory, we need to slightly modify it, 
as will be seen in the next subsection.

\subsection{Monopole configurations}
\label{sec:monopole}

As a configuration with a nontrivial topological charge,
the TP monopole configuration was constructed in 
the $U(2)$ gauge theory on the fuzzy 2-sphere 
\cite{Balachandran:2003ay,AIN3}. 
The TP monopole configuration is given by
\begin{equation}
A_i= L_i^{(n)} \otimes {\bold 1}_2 +
{\bold 1}_{n} \otimes \frac{\tau_i}{2} .
\label{AiLtau}
\end{equation}
where $A_i$'s are the combination (\ref{defAi}).
$L_i^{(n)}$ is the $n$ dimensional representation 
of $SU(2)$ algebra. 
The first and the second factors represent 
the coordinates of the NC space and a configuration 
of the $U(2)$ gauge field respectively. 
The total matrix size is $N=2n$.
The gauge field is given by 
\begin{equation}
a_i=\frac{1}{\rho}{\bold 1}_{n} \otimes \frac{\tau_i}{2}.
\label{monopoleconfig}
\end{equation}
By taking the commutative limit of (\ref{monopoleconfig}), 
and decomposing it into the normal and the tangential components
of the sphere, it can be shown to correspond 
to the TP monopole configuration.
See \cite{AIN3} for details.

Note that $A_i$'s in (\ref{AiLtau}) satisfy the 
$SU(2)$ algebra:
\begin{equation}
[A_i, A_j]=i\epsilon_{ijk}A_k,
\label{SU2alg}
\end{equation}
and can be decomposed 
into irreducible representations: 
\begin{equation}
A_i= U
\begin{pmatrix}
 L_i^{(n+1)} & \cr
& L_i^{(n-1)} \cr
\end{pmatrix}
U^\dag 
\label{decomposition}.
\end{equation}
Hence the TP monopole configuration can be 
geometrically regarded as 
the $2$-concentric fuzzy spheres 
whose matrix sizes differ by two.

More generally, we consider  
$2$-concentric fuzzy spheres 
whose matrix sizes differ by arbitrary integers $2m$:
\begin{equation}
A_i=
\begin{pmatrix}
 L_i^{(n+m)} & \cr
& L_i^{(n-m)} \cr
\end{pmatrix} 
\label{LnpmLnmm}. 
\end{equation}
The $m=0$ case corresponds to the 2-coincident fuzzy 2-spheres,
whose effective action is the $U(2)$ gauge theory
on the fuzzy 2-sphere.
The $|m|=1$ case corresponds to the TP monopole configuration.
The $|m|=n$ case corresponds to the single fuzzy 2-sphere,
whose effective action is the $U(1)$ gauge theory
on the fuzzy 2-sphere.
As we will see in (\ref{monoTCm2}), 
the configuration of $m$ ($|m| \ll n$) corresponds
to the monopole configurations with magnetic charge $-|m|$.

For the configurations of $m\neq 0$, 
$U(2)$ gauge symmetry is broken to 
$U(1)\times U(1)$. 
Thus it is natural to modify the topological 
charge of the rhs in (\ref{indexth}) 
so that it contains only the unbroken 
gauge fields. 
We then define the following modified topological charge:
\begin{equation}
\frac{1}{2}{\cal T}r[\phi'(\Gamma^R +\hat{\Gamma})],
\label{TCunbrokenpro}
\end{equation}
where $\phi'$ is given by 
\begin{equation}
\phi'=\frac{1}{n|m|}\left( A_i^2 -\frac{n^2+m^2-1}{4}\right)
= \frac{m}{2|m|}
\begin{pmatrix}
 {\bold 1}_{n+m} & \cr
& -{\bold 1}_{n-m} \cr
\end{pmatrix} 
.
\label{normalizedphi}
\end{equation}
$\phi' \simeq \frac{\rho}{|m|}\phi $ when $\vert m \vert \ll n$. 
%The last equality holds in the large $n$ limit. 
$\phi'$ is a normalized scalar field in such a sense that 
it satisfies 
%$\phi'^2=1/4$. This corresponds to 
$\sum_a (\phi'^a)^2 =1$, 
if we rewrite 
%$a_i$ as 
$a_i = a_i^a \tau^a /2$ and 
%then $\phi'$ as 
$\phi' = \phi'^a \tau^a /2$ 
%
%This $\phi'$ satisfy  $\sum_{a} (\phi_{a}')^2=1$.
in the commutative limit. 
The modified topological charge (\ref{TCunbrokenpro}) becomes 
\begin{equation} 
\frac{\rho^2}{8\pi}\int_{S^2} d\Omega \epsilon_{ijk}
n_i \phi'^a F_{jk}^a , 
\label{TCunbrokenprocom}
\end{equation}
where $F_{jk}=F_{jk}^a \tau^a/2$ 
is the field strength defined as
$F_{jk}= \partial_j a_k'-\partial_k a_j'-i[a_j',a_k']$,
and $a'_i = {\epsilon_{ijk}x_j a_k / \rho }$. 
This is nothing but the Chern character with the projection 
to pick up the component $\phi'^a$,
namely,
magnetic charge for the unbroken $U(1)$ component
in the TP monopole configuration.

Since $A_i$'s  satisfy $SU(2)$ algebra (\ref{SU2alg}),
we can show by straightforward calculations 
\cite{AIN3} 
\begin{equation}
\frac{1}{2}{\cal T}r[P^{(n \pm m)}(\Gamma^R +\hat{\Gamma})]
=\mp m,
\label{TPpmm}
\end{equation}
for the configuration (\ref{LnpmLnmm}).
Here $P^{(n \pm m)}$ is the projection operator to pick up
the Hilbert space for
the $n \pm m$ dimensional irreducible
representation, respectively.
These projection operators can be written as
\begin{eqnarray}
P^{(n \pm m)}
&=& \frac{(A_i)^2-\frac{1}{4}[(n \mp m)^2-1]}
{\frac{1}{4}[(n \pm m)^2-1]-\frac{1}{4}[(n \mp m)^2-1]} 
\label{P1Ai} \\
&=& \pm \frac{|m|}{m}\phi' + \frac{1}{2}. 
\end{eqnarray}
Hence we can see
\begin{equation}
\frac{1}{2}{\cal T}r[\phi'(\Gamma^R +\hat{\Gamma})]
=-|m|.
\label{monoTCm2}
\end{equation}
Therefore, the configuration (\ref{LnpmLnmm}) represent a 
monopole configuration with a magnetic charge $-|m|$.
In particular, the configuration
(\ref{decomposition}) of $m=1$
is the familiar TP monopole with a magnetic charge $-1$. 
This was also seen  by looking at the explicit form of 
the configuration in (\ref{monopoleconfig}).

Therefore, if by some mechanism,
the $U(2)$ gauge theory,
the 2-coincident fuzzy spheres of $m=0$
decays to the TP monopole configuration of $|m|=1$,
or to the more general monopole configurations of $|m|\ge 1$,
it means the index is generated dynamically.
Also it is understood as a spontaneous symmetry breaking,
$U(2) \rightarrow U(1) \times U(1)$. 
There are two ways to look at the unbroken gauge symmetries. 
\begin{enumerate}
\item Each sphere in (\ref{decomposition}) 
has unbroken $U(1)$ symmetry, and totally $U(1)\times U(1)$. 
This description is geometrically natural.
\item $U(2) \simeq  SU(2) \times U(1)$, and the $SU(2)$ 
breaks down to $U(1)$ when we consider the TP monopole 
configuration (\ref{monopoleconfig}). 
This description becomes natural when we treat the $m=0$ 
configuration as the $U(2)$ gauge theory on the sphere.
\end{enumerate}
These two descriptions are equivalent :
The generators for the $U(1)$ symmetry of each sphere is given by 
\begin{equation}
\begin{pmatrix}
 {\bold 1}_{n+1} & \cr
& 0 \cr
\end{pmatrix}
,
\begin{pmatrix}
 0 & \cr
& {\bold 1}_{n-1} \cr
\end{pmatrix}
.
\end{equation}
They can be rearranged as 
\begin{equation}
\begin{pmatrix}
 {\bold 1}_{n+1} & \cr
& {\bold 1}_{n-1} \cr
\end{pmatrix}
,
\begin{pmatrix}
(n-1){\bold 1}_{n+1}  & \cr
& -(n+1){\bold 1}_{n-1} \cr
\end{pmatrix}
.
\end{equation}
Then, by unitary transformation $U$ 
of (\ref{decomposition}), the second generator becomes
\begin{equation}
U
\begin{pmatrix}
(n-1){\bold 1}_{n+1}  & \cr
& -(n+1){\bold 1}_{n-1} \cr
\end{pmatrix}
U^\dagger=2L_i \tau_i.
\end{equation}
In the commutative limit,
$\alpha L_i \tau_i \rightarrow x_i \tau_i$,
and it is the generator for 
the unbroken $U(1)$ component of the TP monopole.

In order to see the dynamics of 
the index generation through the symmetry breaking,
we will analyze the decay processes from 
the 2-coincident fuzzy spheres
of $m=0$
to configurations of $|m| \ge 1$ in the following section.

\section{Instability of fuzzy spheres}
\setcounter{equation}{0}

In this section we first introduce a 
Yang-Mills-Chern-Simons matrix model,
and then investigate the instabilities of the
monopole configurations in this model. 
We show that the monopole configurations with 
nontrivial topological charges are stabler
than the $U(2)$ gauge theory on the fuzzy 2-sphere.
This realizes a dynamical mechanism of 
the index generation through the symmetry breaking.
The configuration of the 2-coincident fuzzy spheres 
decays to 2-concentric spheres with 
larger matrix size difference,
and cascades into the single fuzzy sphere.

\subsection{The Yang-Mills-Chern-Simons matrix model}

The action of the Yang-Mills-Chern-Simons matrix model
is defined as \cite{Alekseev:2000fd}\cite{IKTW}
\beq
S[A_i] =  
\frac{\alpha ^4}{g^2} 
\, \tr \left( - \frac{1}{4} 
\, [A_{i},A_{j}]^{2} + \frac{2}{3}\, i \, \, \epsilon_{i j k} A_{i} A_{j} A_{k}
   \right) , 
\label{YMCSaction} 
\eeq
where 
$A_i$ ($i = 1, 2 , 3$) are $N\times N$ traceless 
Hermitian matrices
and $\alpha$ is the NC parameter.
This model can be regarded as a dimensionally reduced model 
of $SU(N)$ Yang-Mills theory with the Chern-Simons
term in the three-dimensional Euclidean space. 
It has $SO(3)$ rotational symmetry and $SU(N)$ gauge symmetry
\begin{equation}
A_{i} \to U A_{i} U^{\dag}, 
\end{equation}
where $U \in SU(N)$.

The classical equation of motion is given by 
\begin{equation}
[A_{i},[A_{i},A_{j}]] + i \epsilon_{jkl}[A_{k},A_{l}]=0.
\label{eom}
\end{equation}
The simplest type of solutions is given 
by the commutative diagonal matrices:
\begin{equation}
A_{i}= {\rm Diag}(x_{i}^{(1)}, \cdots, x_{i}^{(N)}).
\end{equation} 
Another type of solution is a fuzzy sphere solution 
which we explained in the previous section and is given by
\begin{equation}
A_i = L_i,
\end{equation}
where $L_i$ ($i = 1,2,3$) are the $SU(2)$ generators. 
We can also consider the following reducible representation:
\begin{eqnarray}
A_i =
\begin{pmatrix}
L_{i} ^{(n_1)} & & & \cr 
& L_{i} ^{(n_2)} & & \cr
& & \ddots & \cr 
& & & L_{i} ^{(n_k)} \cr  
\end{pmatrix}  
\label{general_FSa} ,
\end{eqnarray}
where each
$L_i^{(n)}$ is an $n$-dimensional irreducible representation
of $SU(2)$ algebra,
and thus
$N=\sum_{a=1}^{k} n_a$
is the size of the total matrices.
The case where
$n_1 = \cdots = n_k \equiv n$
corresponds to $k$-coincident fuzzy spheres. 
Expansion of the model around the $k$-coincident fuzzy spheres 
gives $U(k)$ gauge theory
on the fuzzy sphere \cite{IKTW}.

In appendix \ref{oneloop_eff}
we give the calculation of the one-loop effective action in the 
background-field method.

\subsection{Free energies of general monopole configurations}
\label{sec:genmono}
 
We consider the general monopole configurations 
(\ref{LnpmLnmm}).

\begin{figure}[htb]
\begin{center}
\includegraphics[height=8cm]{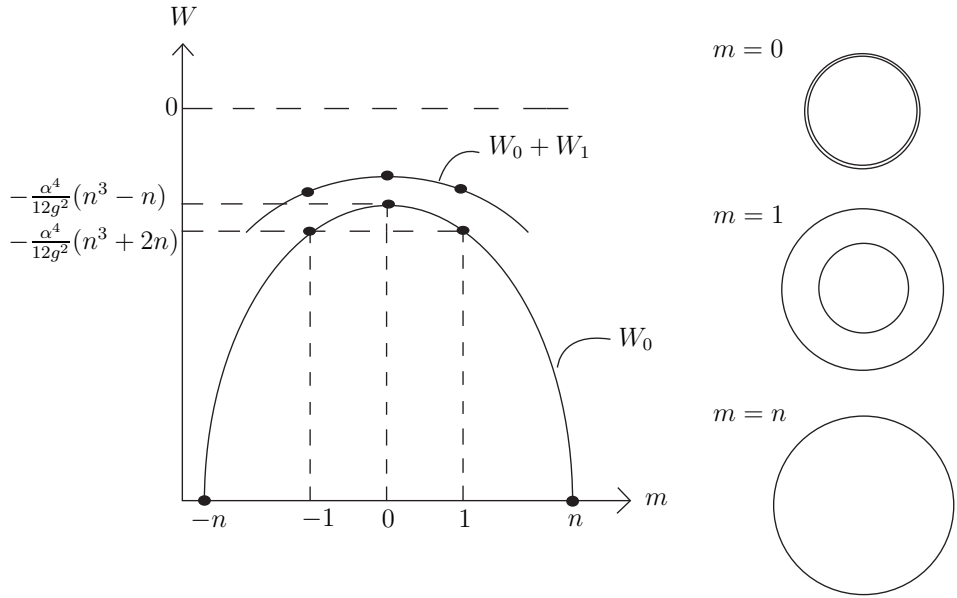}
\end{center}
\caption{The classical action $W_0$ and the one-loop effective 
action $W_1$ as functions of $m$. We also draw the pictures of 
$2$-concentric fuzzy spheres for $m=0$, $m=1$ case and 
a single fuzzy sphere ($m=n$) on the right-side.}
\label{fig:w0w1}
\end{figure}

The classical action for them can be obtained by inserting 
(\ref{LnpmLnmm}) into (\ref{YMCSaction}),
and becomes
\begin{eqnarray}
W_{0} &=& -\frac{\alpha ^4}{24g^{2}} 
              \bigl( \{ (n+m)^3 - (n+m)\} +\{ (n-m)^3 - (n-m) \} \bigr) \nonumber \\
 &=& -\frac{\alpha ^4}{12g^{2}} \bigl( n^3 - n + 3m^2 n \bigr). 
\label{w02con}
\end{eqnarray}
Since it monotonously decreases as $|m|$ increases 
(see Figure \ref{fig:w0w1}),
the  $U(2)$ configuration ($m=0$) is unstable and decays to
the TP monopole configuration ($|m|=1$). 
Repeating such transitions,
it cascades to the $U(1)$ configuration ($|m|= n$).
The processes for $\vert m \vert \ll n$ realize 
the dynamical mechanism of the index generation 
through the symmetry breaking. 
If we consider this mechanism in extra-dimensional spaces 
in some other matrix models, 
chiral fermion in our space-time can be realized
dynamically.
 
We further consider the one-loop correction.
The one-loop effective action can be obtained by inserting 
(\ref{LnpmLnmm}) into (\ref{1loopEA}),\footnote{The $l=0$ 
modes are zero-modes and should be subtracted 
from the one-loop effective action (\ref{onel}). 
Here we write them explicitly in (\ref{onel}) to discuss 
the number of the zero-modes later. }
\begin{eqnarray}
W_{1} &=& \frac{1}{2} {\cal T}r \log 
\bigl( ({\tilde X}_{k})^2 \bigr)  \n \\
&=& \frac{1}{2} \bigl( \sum_{l=0}^{n+m-1} 
+ \sum_{l=0}^{n-m-1}+ 2\sum_{l=m}^{n-1} \ \bigr) 
    (2l+1) \log \left[l(l+1) \right].
\label{onel}
\end{eqnarray}
In the last line the first two terms are contributions from 
the diagonal blocks while the last term comes from the 
off-diagonal blocks.
Since ${\tilde X}_{i} M = (L_i^{(n+m)L}-L_i^{(n-m)R})M$
when ${\tilde X}_{i}$ acts on the upper-right off-diagonal 
block $M$,
its representation can be obtained by adding two 
$SU(2)$ representations with dimension $(n+m)$ and $(n-m)$.

We can see from (\ref{onel}) that 
the configuration (\ref{LnpmLnmm}) 
of $m=0$ has 4 zero-modes,
while those of $m \neq 0$ have 2.
We will discuss the instability in these zero-mode directions
for $m=0$ case in subsection \ref{sec:ana2coincident}
and for $m=1$ case in subsection \ref{sec:anaTP}.
In this subsection we will first consider the contributions
from the non-zero-modes.

For large $n$, the summation over $l$ in (\ref{onel}) 
can be replaced by an integration.
Thus, for $m=0$,
\begin{equation}
W_{1}^{m=0} =
2 \bigl[ n(n-1)\bigl(\log n(n-1) -1\bigr)
- 2\bigl(\log 2 -1\bigr)\bigr] ,
\label{wim0}
\end{equation}
and for $m \neq 0$, 
\begin{eqnarray}
W_{1}^{m\neq 0}   
&=& \frac{1}{2} \left[ (n+m)(n+m-1)\bigl(
\log [(n+m)(n+m-1)] 
-1\bigr) \right. \nonumber \\
& &                   \ \ \ 
+ (n-m)(n-m-1)\left( \log 
[(n-m)(n-m-1)] -1 \right) \nonumber \\
& &                   \ \ \ 
   + 2n(n-1)\bigl(\log n(n-1) -1\bigr)  \nonumber \\
&&     \ \ \
-2\sum_{l=1}^{m-1} (2l+1) \log \left[l(l+1) \right] \nonumber \\  
& &            \ \ \  
\left. -4 \cdot 2\bigl(\log 2 -1\bigr) \right]. 
\end{eqnarray}
For $m \ll n$, the difference between them 
can be evaluated as  
\begin{eqnarray}
\Delta W_{1} &=& W_{1}^{m \neq 0} -W_{1}^{m=0} \n \\            
&=& \frac{1}{2}[f(n\!\!+\!\!m) +f(n\!\!-\!\!m) -2f(n)]
 -\sum_{l=1}^{m-1} (2l+1) \log \left[l(l+1) \right] \n \\
&\simeq & \frac{1}{2}[ f^{(2)}(n)m^2 +\frac{1}{12}f^{(4)}(n)m^4 
+\cdots] -\sum_{l=1}^{m-1} (2l+1) \log \left[l(l+1) \right] \n \\                       
&=& \frac{1}{2} \left[ 2\log n(n-1)+ 
\frac{(2n-1)^2}{n(n-1)} \right] m^2 
+ {\cal O}(1/n^2)m^4 +\cdots \nonumber \\
& & -\sum_{l=1}^{m-1} (2l+1) \log \left[l(l+1) \right] \n \\
&\simeq& 2m^2 \log n,
\label{deltaw1}
\end{eqnarray}
where in the second line we introduced 
$f(x) = x(x-1)\{ \log x(x-1)-1\}$,
and in the last line we took the leading term in the large $n$ limit.

From (\ref{wim0}) and (\ref{deltaw1}),
at large $n$,
$m$-independent term in $W_1$ is of the order of $n^2 \log n$
and $m$-dependent term in $W_1$ is of the order of $(\log n) m^2$,
while from (\ref{w02con}),
$m$-independent term in $W_0$ is of the order of $n^3$
and $m$-dependent term in $W_0$ is of the order of $n m^2$:
\begin{eqnarray}
W_0 &\sim& -n^3 - n m^2,  \\ 
W_1 &\sim& n^2 \log n + (\log n) m^2. 
\end{eqnarray}
Therefore the vacuum structure mentioned after 
(\ref{w02con}) does not change qualitatively at large $n$ 
even after we take into consideration 
the quantum corrections. 
Namely, the vacuum structure is determined classically. 
The classical action $W_0$ and the effective action with the 
one-loop correction are depicted in Figure \ref{fig:w0w1} 
as a function of $m$.

%%%%%%%%%%%%%%%%%%%%%%%%%%%%%%%%%%%%%%%%%%%%%%%%

\subsection{Instability of the $U(2)$ gauge theory}
\label{sec:ana2coincident}

In this subsection we analyze the 
instability of the $U(2)$ gauge theory,
the 2-coincident fuzzy spheres of $m=0$, along the 
zero-modes explained in the previous subsection.
We also consider the decay process from
this configuration to the TP monopole configuration
of $|m|=1$.

In order to see 
how the 2-coincident fuzzy spheres ($m=0$) decay, 
we consider the zero-mode directions
around this configuration.
As we mentioned after (\ref{onel})
there are 4 zero-modes, 
one of which is $A_i \sim {\bold 1}_{2n}$,
the total translation,
and should be neglected due to the tracelessness condition 
imposed on the matrices $A_i$.  
We thus consider the following background
\footnote{The background (\ref{tau_dir}) with 
only $a=3$ direction for $h_i^a$ was considered 
in Appendix D in \cite{Azuma:2004zq}. 
This direction corresponds to the one in which
the positions of two fuzzy spheres shift relatively.
However, the 2-coincident spheres
seem to decay to the TP monopole configuration
by changing the sizes of the two spheres,
as can be seen from the analysis of subsection \ref{sec:genmono}
and also from the numerical analysis
in \cite{Azuma:2004zq}.
Thus it will be better to consider 
the direction (\ref{decaypath}) 
to discuss the decay process.
}
\begin{equation}
X_i= L_i^{(n)} \otimes {\bold 1}_2 +
h_i^a {\bold 1}_{n} \otimes \tau^a. 
\label{tau_dir} 
\end{equation}

The classical action can be obtained by 
inserting (\ref{tau_dir}) into (\ref{YMCSaction}), 
and becomes
\begin{eqnarray}
W_0 &=& -\frac{\alpha ^4}{12g^{2}} \bigl( n^3 - n \bigr) \n \\
& & 
+ \frac{\alpha ^4}{g^{2}} n 
\left[ 2\{ (h_i^a h_i^a)^2 - (h_i^a h_j^a)^2 \} 
 - 8\det (h_i^a)\right],
\label{2coin_cl}
\end{eqnarray}
which has the third and fourth order terms in $h_i^a$.
Since the third order term $-8\det(h_i^a)$ becomes minimum
in the direction of $h_i^a \propto \delta_{ia}$,
the two-coincident fuzzy spheres decay into this direction.
We also show in appendix \ref{sec:absmin}
that (\ref{2coin_cl}) takes an absolute minimum
at $h_i^a=\frac{1}{2}\delta_{ia}$,
which is nothing but the TP monopole configuration.
Therefore we infer that the 2-coincident fuzzy spheres decay into
the TP monopole configuration along the path
\begin{equation}
X_i= L_i^{(n)} \otimes {\bold 1}_2 +
h{\bold 1}_{n} \otimes \frac{\tau^i}{2} 
\label{decaypath}
\end{equation}
from $h=0$ to $h=1$.
 
Next we evaluate the one-loop correction 
around the background (\ref{tau_dir}).
The one-loop effective action can be obtained by
inserting (\ref{tau_dir}) into (\ref{1loopEA}),
\begin{eqnarray}
W_{1} &=& \frac{1}{2} {\cal T}r \ \tr' 
\log \left[ ({\tilde X}_{k})^2 \delta_{ij} 
- 2i \epsilon_{ijk}h_k^a {\tilde \tau}^a 
+ 4i h_i^a h_j^b \epsilon_{abc} {\tilde \tau}^c \right] \n \\
& & - {\cal T}r \log \left[ ({\tilde X}_{k})^2 \right], 
\end{eqnarray}
where ${\tilde X}_i$ and ${\tilde \tau}_i$ are adjoint operators 
which act on a hermitian matrix $M$ as 
\begin{eqnarray}
{\tilde X}_i M
&=&[L_i^{(n)} \otimes {\bold 1}_2 +
h_i^a {\bold 1}_{n} \otimes \tau^a , M],\\
{\tilde \tau}_i M&=&[\tau_i,M].
\end{eqnarray}
Up to the second order of the perturbative expansion in $h_i^a$, 
the one-loop effective action becomes 
\begin{eqnarray} 
W_1 &=& 2 \sum_{l=1}^{n-1} (2l+1) \log \left[l(l+1) \right] \n\\
& & + (h_i^a)^2 
\left[\frac{4}{3} \sum_{l=1}^{n-1}\frac{2l+1}{l(l+1)}
-16 \sum_{l=1}^{n-1}\frac{2l+1}{(l(l+1))^2}  \right],
\end{eqnarray}
where we did not include the zero-modes ($l=0$) 
since they are collective 
modes and should be treated separately.
The coefficient of $(h_i^a)^2$ 
changes its sign from negative to positive at $n=374$,
and at large $n$ limit,
\begin{eqnarray}
W_1
&\simeq& 
2 \bigl[ n(n-1)\bigl(\log n(n-1) -1\bigr)
-2(\log 2 -1) \bigr] \n \\
& & + (h_i^a)^2 \left[ \frac{4}{3}\bigl(\log n(n-1) 
- \log 2 \bigr)  -16 \left( 1-\frac{1}{n^2} \right) \right].
\end{eqnarray}

\begin{figure}[htbp]
\begin{center}
\subfigure[The classical potential $V_0$]
{\includegraphics[width=6.5cm]{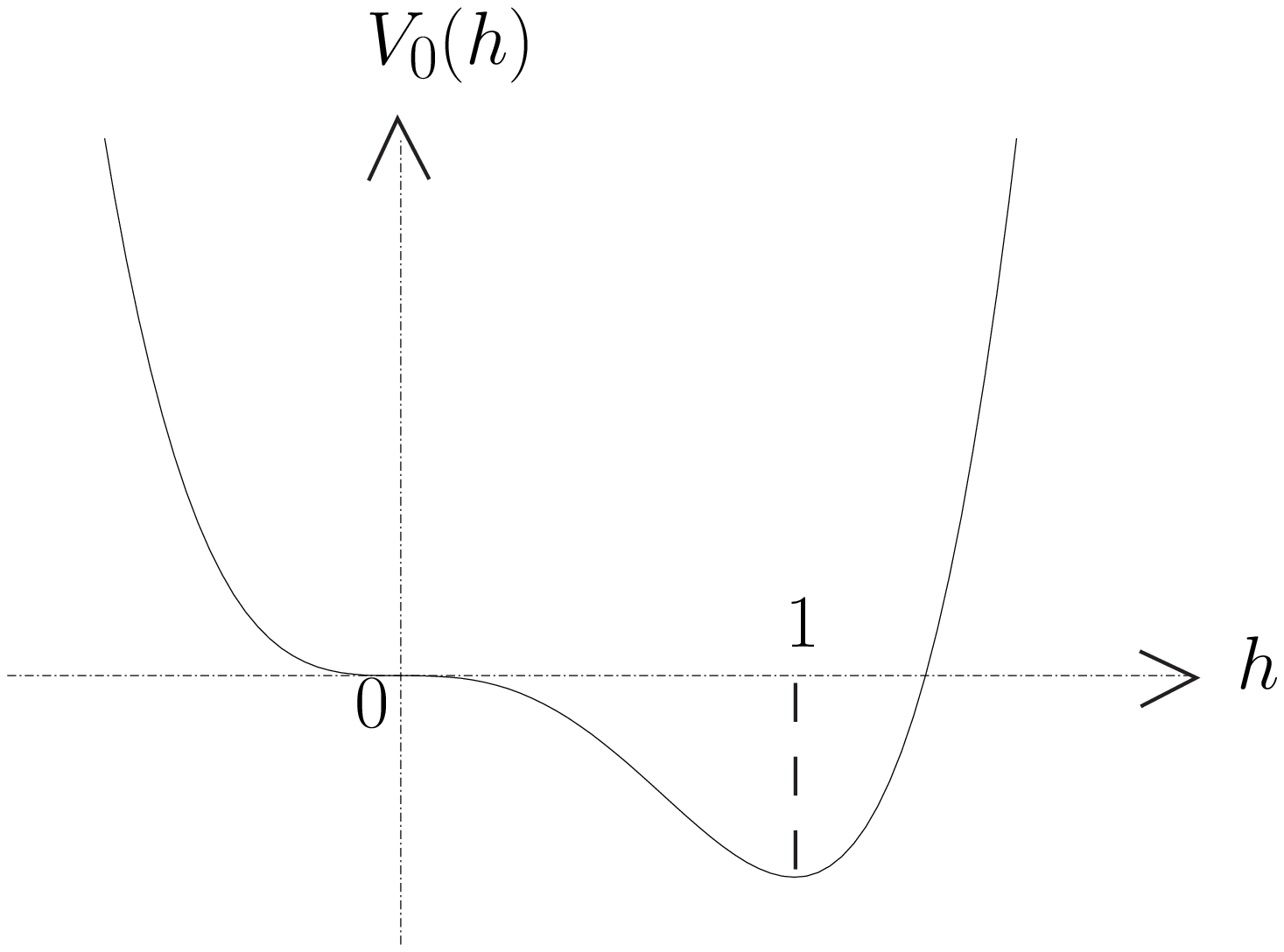}}
\label{classicalv1}
\\
\subfigure[$V_0+V_1$ for small $n$]
{\includegraphics[width=6.5cm]{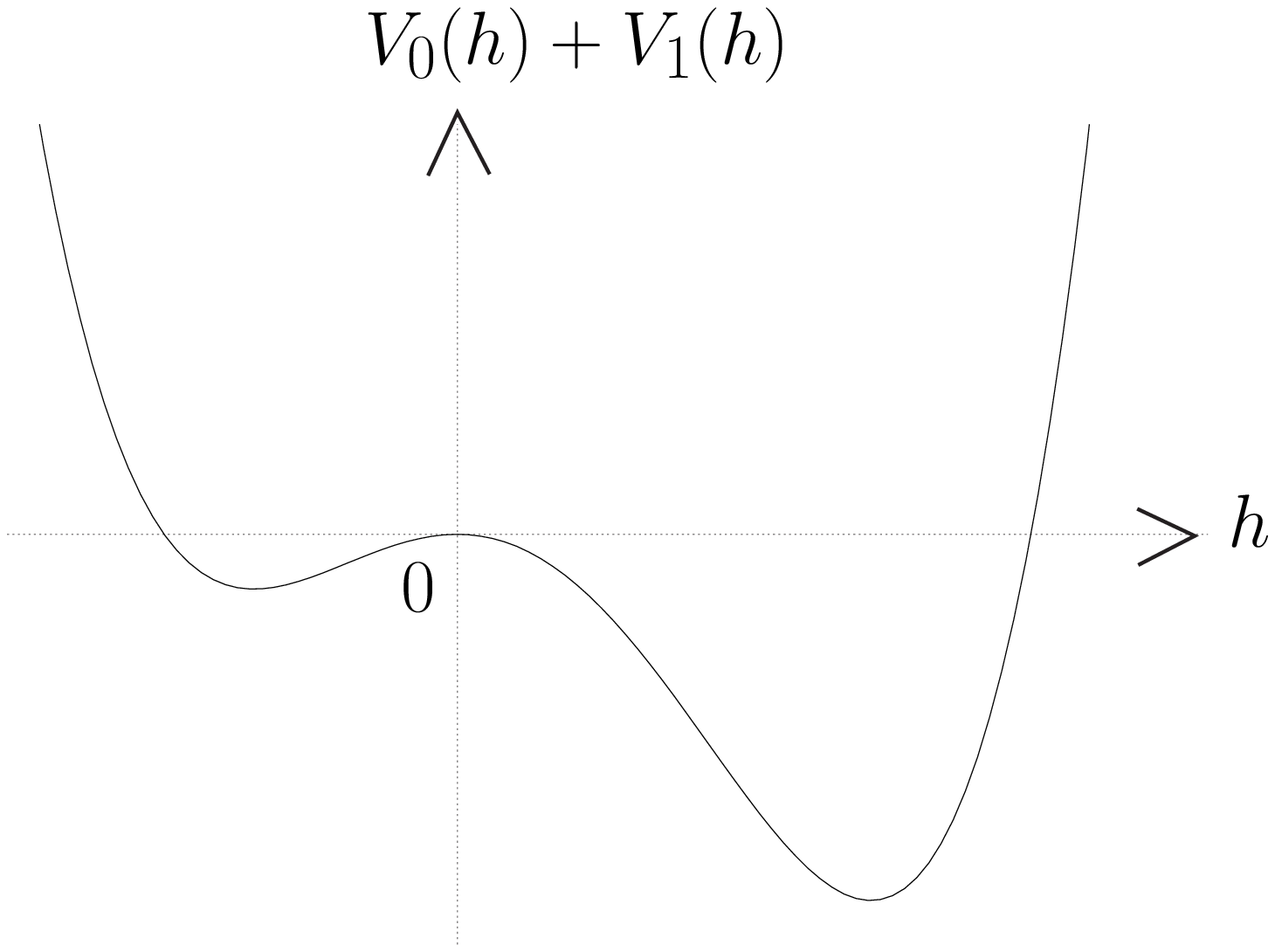}}
\label{smalln} 
\subfigure[$V_0+V_1$ for large $n$]
{\includegraphics[width=6.5cm]{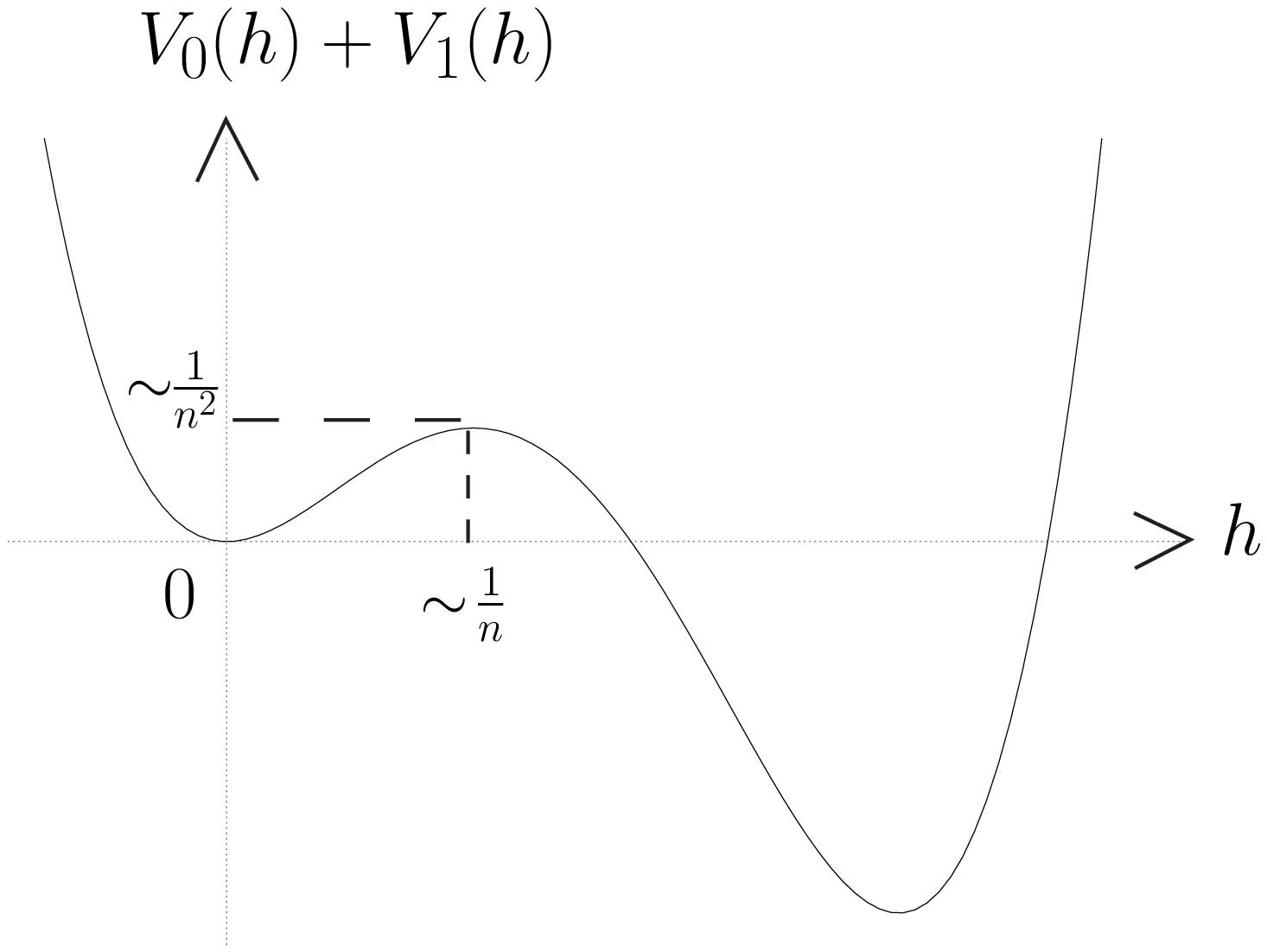}}\label{largen}
\end{center}
\caption{(a)The classical potential is flat at $h=0$ and 
takes a minimum value at $h=1$.
(b)The one-loop potential is unstable at $h=0$ for small values 
of $n$.
(c)The one-loop potential is metastable at $h=0$ for large 
values of $n$.}
\label{fig:epcoin}
\end{figure}

In summary,
the configuration of 2-coincident fuzzy spheres  
has 3 nontrivial zero-modes,
which include the decay direction
as was seen from the classical action (\ref{2coin_cl}).
Including the one-loop correction,
all of the 3 zero-mode directions become unstable for 
$n \leq 373$, and stable for $n \ge 374$.
Thus this configuration becomes metastable
for large $n$.
Then the transition to the TP monopole configuration
must be qualitatively different.
However, 
since the one-loop contribution is like $(\log n) h^2$
while the classical contribution is like $-n h^3$,
the metastability becomes negligible at large $n$ limit.
See Figure \ref{fig:epcoin}.

We will illustrate this feature by plotting $W_0+W_1$
along the path (\ref{decaypath}). 
\begin{eqnarray}
W_0 + W_1 
&=&
-\frac{\alpha ^4}{12g^{2}} \bigl( n^3 - n \bigr)  
+2 \sum_{l=1}^{n-1} (2l+1) \log \left[l(l+1) \right]  \n \\
&&
+V_0(h)+V_1(h) ,  
\end{eqnarray}
where 
\begin{eqnarray}
V_0(h) &=&  \frac{\alpha ^4 n}{g^{2}} 
\Bigl( \frac{3}{4}h^4 - h^3 \Bigr) ,
\label{v0hdecay}\\
V_1 (h) &=&
 \biggl[\sum_{l=1}^{n-1}\frac{2l+1}{l(l+1)}
 - \! 12\Bigl(1 \!- \! \frac{1}{n^2}\Bigr)\!\biggr] h^2 .
\end{eqnarray} 
In Figure \ref{fig:epcoin} 
we plot the classical potential $V_0(h)$ in (a), 
the one-loop effective potential $V_0(h)+V_1(h)$ 
for small values of $n$ in (b), 
and for large values of $n$ in (c). 

\subsection{Instability of the TP monopole configuration}
\label{sec:anaTP}
In the previous subsection we analyzed the transition
from the $U(2)$ gauge theory to the TP monopole configuration.
As we saw in subsection \ref{sec:genmono}, 
the TP monopole configuration is not stable
and decays into the configuration of larger $|m|$.
We analyze the instability 
of the TP monopole in this subsection.
To see how
the TP monopole configuration decays, 
we consider the zero-mode directions around this configuration.
As we mentioned after (\ref{onel})
there are 2 zero-modes, 
one of which is $A_i \sim {\bold 1}_{2n}$,
the total translation,
and should be neglected.
Thus we consider the following background
\begin{equation}
X_i= \left(
\begin{array}{cc}
L_i^{(n+1)} & \\
& L_i^{(n-1)} 
\end{array} \right)
+h_i
\left(
\begin{array}{cc}
\frac{n-1}{n} {\bold 1}_{n+1} & \\
& -\frac{n+1}{n} {\bold 1}_{n-1}
\end{array}\right)
.
\label{TPfluc} 
\end{equation}

The classical action is obtained by inserting (\ref{TPfluc})
into (\ref{YMCSaction}), and becomes 
\begin{equation}
W_{0} = -\frac{\alpha ^4}{12g^{2}} \bigl( n^3 +2 n \bigr) .
\end{equation}
Note that this direction is flat not only 
in the second order but in all orders in $h_i$.
Thus we have to find the direction along which the 
TP monopole configuration decays, including the non-zero-modes.
It is an interesting study to find such a direction 
that the potential along the path $h$ has the form like
\begin{equation}
V_0(h) \sim n(h^2-h^3),
\label{TPmetastacla}
\end{equation}
and the path connects the TP monopole configuration ($|m|=1$)
to the configuration of $|m|=2$. 
In any case, the TP monopole configuration must be metastable
classically.

We now evaluate the one-loop effective action around the 
background (\ref{TPfluc}).
By inserting it into (\ref{1loopEA}), we obtain
\begin{eqnarray}
W_{1} &=& \frac{1}{2} {\cal T}r \ \tr' 
\log \bigl[ ({\tilde X}_{k})^2 \delta_{ij} 
- 2i \epsilon_{ijk} {\tilde H}_k \bigr] \nonumber \\
& & - {\cal T}r \log \bigl( ({\tilde X}_{k})^2 \bigr), 
\end{eqnarray}
where ${\tilde X}_i$ and ${\tilde H}_i$ are adjoint operators 
which act on a hermitian matrix $M$ as 
\begin{eqnarray}
{\tilde X}_i M &=& 
\Biggl[ \biggl(
\begin{array}{cc}
L_i^{(n+1)} & \\
& L_i^{(n-1)} 
\end{array} 
\biggr)
+h_i
\biggl(
\begin{array}{cc}
\frac{n-1}{n} {\bold 1}_{n+1} & \\
& -\frac{n+1}{n} {\bold 1}_{n-1}
\end{array}
\biggr)
, M \Biggr], \\
{\tilde H}_i M
&=&\Biggl[ h_i
\biggl( 
\begin{array}{cc}
\frac{n-1}{n}{\bold 1}_{n+1} & \\
& -\frac{n+1}{n} {\bold 1}_{n-1}
\end{array} 
\biggr)  ,  M \Biggr] . 
\end{eqnarray}
Up to the second order of the perturbation in $h_i$,
\begin{eqnarray}
W_1
&=&\frac{1}{2} \biggl( \sum_{l=1}^{n} 
+\sum_{l=1}^{n-2} +2\sum_{l=1}^{n-1} \biggr) 
    (2l+1) \log \left[ \,  l(l+1) \right] \nonumber \\
& & + (h_i)^2 
\Biggl[ \frac{4}{3} \sum_{l=1}^{n-1}\frac{2l+1}{l(l+1)} 
-16 \sum_{l=1}^{n-1} \frac{2l+1}{(l(l+1))^2} \Biggr],
\end{eqnarray}
where again we have subtracted the zero-modes 
since they are collective 
modes and should be treated separately.
The coefficient of $(h_i)^2$ 
changes its sign from negative to positive at $n=374$,
and becomes $8/3 \log n$ at large $n$ limit.
Therefore, along this direction, the TP monopole configuration
is unstable for small $n$, while stable for large $n$.

As for the decay from the TP configuration,
we must consider other directions.
As we mentioned in (\ref{TPmetastacla}),
the TP configuration is metastable even at the classical level. 
The quantum corrections of the order of $(\log n) h^2$
will not change the qualitative feature of the classical 
metastability. 
It is plausible that all of the monopole configurations of 
$|m| \ge 1$ have the same property,
since they also have only one nontrivial zero-mode.

\section{Conclusion and Discussion}
\setcounter{equation}{0}

In this paper
we presented a dynamical mechanism for 
an index generation
through the spontaneous symmetry breaking: 
$U(2)\simeq SU(2)\times U(1) \rightarrow U(1)\times U(1)$,
by showing that 
the monopole configurations with the nontrivial 
topological charges are stabler
than the $U(2)$ gauge theory without any condensation 
on the fuzzy 2-sphere, 
%%%%%%%%%%%%%%%%%%%%%%%%%%%%%%%%%%
though it finally decays to a single fuzzy 2-sphere 
whose effective theory is the $U(1)$ gauge theory 
on the fuzzy 2-sphere. 
%which is expressed in terms of irreducible representation of 
%$SU(2)$ in the present model. 
The final state is geometrically 
different from the initial state, two-coincident fuzzy spheres. 
Nevertheless, we expect that 
such a mechanism of the dynamical generation of an index 
as studied in this paper would be useful to 
realize chiral fermions dynamically 
by compactifying extra-dimensional spaces 
with a nontrivial index in some more realistic matrix models.
%%%%%%%%%%%%%%%%%%%%%%%%%%%%%%%%%%%%
%Then, if we use this mechanism in extra-dimensional space,
%chiral fermion in our space-time can be realized
%dynamically.

We first analyzed the instability of 
the general monopole configurations (\ref{LnpmLnmm})
in the Yang-Mills-Chern-Simons matrix model.
The classical action for these configurations
monotonously decreases as the difference of the
sizes of the two matrices, $|m|$, increases 
(see Figure \ref{fig:w0w1}).
Thus
the $U(2)$ gauge theory, 
the 2-coincident spheres of $m=0$, 
is unstable and decays to
the TP monopole configuration of $|m|=1$. 
Repeating such transitions,
it cascades to 
the $U(1)$ gauge theory,
the single sphere of $|m|= n$.
These properties do not change
even after we take into consideration the quantum corrections.
Namely, the vacuum structure is determined classically.

We then analyzed the instability of 
the $U(2)$ gauge theory ($m=0$),
in detail.
This configuration has 3 nontrivial zero-modes,
which includes the direction along which it 
decays to the TP monopole ($|m|=1$).
Including the one-loop correction,
all of the 3 zero-mode directions become unstable for $n \le 373$,
and stable for $n \ge 374$.
Thus the $U(2)$ gauge theory  becomes metastable
for large $n$.
Then the transition to the TP monopole configuration
must be qualitatively different:
2nd order like transition for small $n$
and 1st order like for large $n$.  
However, 
since the one-loop contribution is like $(\log n) h^2$
while the classical contribution is like $-n h^3$,
the metastability becomes negligible at large $n$ limit.

It will be interesting to 
study the decay processes
from the $U(2)$ gauge theory to the TP monopole configuration
in detail.
If we introduce an extra time direction and consider the 
action (\ref{YMCSaction}) as the potential, we can discuss 
the instability of the fuzzy sphere 
in the M(atrix)-theory\cite{Banks:1996vh} 
or the instability observed in the Monte Carlo simulation 
of the matrix model in \cite{Azuma:2004zq}.
Then we can evaluate the decay rate
by using the path (\ref{decaypath}) 
in some semiclassical method. 

We further analyzed the instability of the 
TP monopole configuration ($|m|=1$).
This configuration has one nontrivial zero-mode,
which is a flat direction classically.
The one-loop quantum correction makes it
unstable for small $n$
and stable for large $n$.
As for the decay 
we have to consider other directions,
which include the non-zero-modes. 
Thus the TP monopole configuration must be 
metastable even at the classical level,
and quantum corrections will not change this property. 
It is plausible that all of the configurations of 
$|m| \ge 1$ have this property
since they have only one zero-mode.

It is interesting to find the decay path from
the TP configuration of $|m|=1$ to the one of $|m|=2$,
and the path from $|m|=2$ to $|m|=3$,
and so on,
and check the above mentioned conjecture.
It is also interesting to clarify 
the explicit form of the configuration of 
the general monopoles of $|m| \ge 2$, 
as we did for $|m|=1$ in \cite{AIN3}.

Another issue is to study topologically nontrivial configurations
in other NC geometries than the fuzzy 2-sphere 
and present a mechanism for the
dynamical generation of an index
on general NC geometries.
We might be able to use this mechanism
in the matrix models for 
the critical string theories like \cite{IKKT},
and realize the chiral gauge theory 
in our four-dimensional space-time.

\section*{Acknowledgements}
We would like to thank T. Azuma, S. Bal, Y. Kitazawa,
B. Morariu
and J. Nishimura for discussions and useful comments.
We are also grateful to the referee 
for giving us valuable comments to improve our first manuscript.

\appendix

%%%%%%%%%%%%%%%%%%%%%%%%%%%%%%%%%%%%%%%%%%%%%%%%
\section{The one-loop effective action}\label{oneloop_eff}
\setcounter{equation}{0}

Here we give the calculation of the one-loop effective action 
in the background-field method.
We consider backgrounds $X_i$ 
and fluctuations around it, $\tilde{A}_{i}$: 
\begin{equation}
A_{i} =X_{i}+\tilde{A}_{i}. \label{dec}
\end{equation}
We add to the action (\ref{YMCSaction}) the gauge fixing term and the ghost term:
\begin{eqnarray}
S_{\rm g.f.} &=& -\frac{1}{g^{2}} \tr [X_i, A_i]^2 \ , \\
S_{\rm ghost}&=& -\frac{1}{g^{2}} \tr  \left([X_i,\bar{c}][A_i,c] \right) \ ,
\end{eqnarray}
where $c$ and $\bar c$ are the ghost and anti-ghost fields, respectively.

We then expand the action (\ref{YMCSaction}) 
up to the second order in the fluctuations:
\begin{eqnarray}
S_{\rm total} &=& S[X_i]+ S_{2}  \ ,\\ 
S_{2} &=& \frac{1}{2g^{2}} \tr \left( 
\tilde A_i [X_k , [X_k , \tilde A_i ]] 
- 2\bigl( [X_i , X_j] - i \epsilon_{ijk} X_k \bigr)
[\tilde A_i , \tilde A_j ] \right) \nonumber \\ 
& & + \frac{1}{g^{2}} 
\tr \Bigl( \bar c \, [X_k , [X_k , c ]] \Bigr) \n \\ 
&=& \frac{1}{2g^{2}} \tr \left( 
\tilde A_i \bigl[ ({\tilde X}_{k})^2 \delta _{ij} 
+ 2\bigl([{\tilde X}_i , {\tilde X}_j]
-i \epsilon_{ijk}  {\tilde X}_k \bigr)\bigr] \tilde A_j \right) \nonumber \\ 
& & + \frac{1}{g^{2}} \tr 
\Bigl( \bar c \ ({\tilde X}_{k})^2 c \Bigr),
\end{eqnarray}
where in the last step we introduced adjoint operators, 
\begin{eqnarray}
{\tilde X}_{i} M &=& [X_{i}, M],\\
\bigl( [{\tilde X}_i , {\tilde X}_j]-i 
\epsilon_{ijk} {\tilde X}_k \bigr)M
&=&
[\bigl( [X_i , X_j]-i \epsilon_{ijk} X_k \bigr),M].
\end{eqnarray}
Note that we drop the linear terms in $\tilde A_i$ by hand
in the background-field method.

The one-loop effective action $W_1$ is obtained as
\begin{eqnarray}
W_{1}
&=&-\log \int d\tilde{A} dc d\bar c \ e^{-S_{2}} \n \\
&=& \frac{1}{2} {\cal T}r \  \tr' \log \bigl[ 
({\tilde X}_{k})^2 \delta _{ij} 
+ 2\bigl( [{\tilde X}_i , {\tilde X}_j] 
- i \alpha \epsilon_{ijk} {\tilde X}_k \bigr) \bigr] \nonumber \\ 
& & - {\cal T}r \log \bigl[ ({\tilde X}_{k})^2 \bigr],
\label{1loopEA}
\end{eqnarray}
where ${\cal T}r$ is a trace of operators acting on matrices, 
and $\tr'$ is the trace over the space-time indices $i$ and $j$.

\section{Analysis of the potential (\ref{2coin_cl})}
\label{sec:absmin}
\setcounter{equation}{0}

In this appendix we show that the potential (\ref{2coin_cl})
has the absolute minimum at $h_i^a = \frac{1}{2}\delta_{ia}$,
which is nothing but the TP monopole configuration.
The potential (\ref{2coin_cl}) can be written as 
\begin{equation}
V (H) = 2[(\tr H^T H)^2 - \tr (H^T H H^T H) ] - 8\det (H),
\end{equation} 
where $(H)_{ia}= h_i^a$.
Introducing $ M = H^T H$, 
\begin{eqnarray}
  V(H)   &=& 2[ (\tr M)^2 - \tr (M^2) ] \mp 8\sqrt{\det M} \n \\
     &=& 2\Bigl[ \Bigl(\sum\nolimits_{i=1}^{3} x_i\Bigr)^2 
         -\sum\nolimits_{i=1}^{3} x_i^2 \Bigr] \mp 
         8\sqrt{ \prod_{i}x_i } \n \\
     &=& 4(x_1 x_2 + x_2 x_3 + x_3 x_1 ) 
\mp 8\sqrt{x_1 x_2 x_3} \ ,
\label{VH}
\end{eqnarray}
up to the overall coefficient.
Here $x_1,x_2,x_3$ are the eigenvalues  of $M$.
Fixing the value of $x_1 x_2 x_3$, 
the first term in eq.(\ref{VH}) has a lower bound as 
\begin{eqnarray}
x_1 x_2 + x_2 x_3 + x_3 x_1 
&=& (x_1 x_2  x_3)
\left( \frac{1}{x_1}+\frac{1}{x_2}+\frac{1}{x_3}\right) \n \\
&\geq& 3(x_1 x_2  x_3) 
\left( \frac{1}{x_1 x_2 x_3}\right)^{\frac{1}{3}} \n \\
&=& 3(x_1 x_2  x_3)^{\frac{2}{3}}, 
\end{eqnarray}
where in the second inequality the equality is satisfied when
$x_1 = x_2 = x_3 \equiv y^2$.
In this case $V(H)$ is given by 
\begin{eqnarray}
V(H) &=&  12y^4 -8y^3  \n \\
&\geq& -\frac{1}{4},  
\end{eqnarray}
where in the second line equality is satisfied when $y=\frac{1}{2}$.

Therefore the minimal point $V(H)=-1/4$
is satisfied at $h_i^a =\frac{1}{2}\delta_i^a$
up to $SU(2)$ rotation.
Thus $W_0$ in (\ref{2coin_cl})
is minimized at the TP monopole configuration
under the restriction on possible directions to move in the 
configuration space as in eq(\ref{tau_dir}).

\end{document}